\documentclass[prd,showpacs,preprintnumbers,nofootinbib,eqsecnum]{revtex4}

\usepackage[dvipdfm]{graphicx}
\usepackage{latexsym}
\usepackage{amsfonts}
\usepackage{amssymb}
\usepackage{amsmath}
\usepackage[usenames]{color}

\begin{document}

\preprint{YITP-12-11, KUNS-2386}

\title{Destabilizing Tachyonic Vacua at or above the BF Bound}
\author{Sugumi Kanno$^{1}$}
\author{Misao Sasaki$^2$}
\author{Jiro Soda$^{3}$}

\affiliation{$^{1}$Institute of Cosmology,
Department of Physics and Astronomy, 
Tufts University, Medford, Massachusetts 02155, USA \\
$^2$ Yukawa Institute for Theoretical Physics, Kyoto University,
Kyoto 606-8502, Japan\\
$^{3}$Department of Physics,  Kyoto University, Kyoto 606-8502, Japan\\
}%

\date{\today}

\begin{abstract}
It is well known that tachyonic vacua in an asymptotically
 Anti-de Sitter (AdS) spacetime
are classically stable if the mass squared 
is at or above the Breitenlohner and Freedman (BF) bound.
We study the quantum stability of these tachyonic vacua in
 terms of instantons.
We find a series of exact instanton solutions destabilizing
 tachyonic state at or above the BF bound in asymptotically AdS space. 
We also give an analytic formula for the decay rate and show that
it is finite.
Comparing our result with
the well-known algebraic condition for the stability,
we discuss stability conditions of tachyonic vacua at or 
above the BF bound. 
\end{abstract}

\maketitle
\section{Introduction}

In the Minkowski background, the classical stability of a vacuum
of a scalar field is guaranteed if its mass squared 
(ie the curvature of the potential) is positive.
However, this does not imply the quantum stability of the vacuum.
It may be metastable depending on the global shape of the potential.
If the vacuum is a local minimum but not a global minimum of the 
potential, it may decay into a stabler state via the bubble
nucleation due to quantum tunneling~\cite{Coleman:1977py,Callan:1977pt}. 
This situation remains essentially the same even if we take gravity
into account~\cite{Coleman:1980aw}. 

In an AdS background, the situation is more complicated.
Indeed, it is known that an AdS vacuum may be classically stable
even if the mass squared is negative, provided that it is at or 
above the BF bound~\cite{Breitenlohner:1982jf,Breitenlohner:1982bm},
\begin{eqnarray}
m^2_{\rm BF} \ell^2 = - \frac{(d-1)^2}{4} \ ,
\label{bf}
\end{eqnarray}
where $\ell$ is the curvature radius of the AdS spacetime
and $d$ is the spacetime dimension.
Of course, as in the case of the Minkowski spacetime,
this does not guarantee the quantum stability of the system.
However, in contrast to the Minkowski case, one cannot
judge the stability of a tachyonic vacuum just by looking 
at the global shape of the potential.
In fact, the system is perfectly stable even
for an unbounded potential such as an inverted quadratic 
potential provided that the mass squared satisfies
the BF stability bound.
So, the question is how to judge the stability of a vacuum. 

To this end, we mention that there exists an algebraic criterion
 for the global stability of an AdS vacuum.
It was proved that if the potential takes the form,
\begin{eqnarray}
  V(\phi) = (d-2) \left(\frac{dP(\phi)}{d\phi}\right)^2
 - (d-1) P(\phi)^2\,,
\label{super}
\end{eqnarray}
then any non-trivial field configuration has
a positive definite energy~\cite{Boucher:1984yx,Townsend:1984iu}.
Hence, the system is stable. 
In other words, considering (\ref{super}) as a first-order
nonlinear differential equation for $P$ for a given potential $V$,
the existence of a real solution for $P$ guarantees
the absolute stability of the system.
This is a sufficient condition for
 the stability (The stability condition is further refined recently in
\cite{Hertog:2005hm,Amsel:2006uf,Amsel:2007im,Faulkner:2010fh,Amsel:2011km}).
Even for this sufficient condition, knowing the explicit form
of a potential does not help us much to know the stability
since it involves solving the nonlinear differential equation. 

Given this situation, we take another strategy. 
Namely, we consider instability conditions instead of stability
conditions.
 It is known that if there exists an instanton in the system,
then it is quantum mechanically unstable.
Thus we look for instanton solutions. Specifically, we
look for a class of potentials which admit exact analytical 
instanton solutions.

To the best of our knowledge, no one has found instantons 
representing the decay of a tachyonic vacuum
(except for instantons with unconventional boundary
conditions~\cite{Hertog:2004rz,Hertog:2005hu}
which are not related to the vacuum decay).
This means no one has found a potential which is classically stable
 but quantum mechanically unstable.

In this paper, extending the method developed
in \cite{Kanno:2011vm,Kanno:2012zf}, we look for
exact instantons for a class of 
potentials which satisfies the BF stability bound.
The class of potentials contain two parameters; one
of them controls the value of the mass squared and
the other the global shape of the potential. 
We compute everything analytically. In particular,
the bounce action is analytically computed and found
to be positive and finite. 

Thus we find a class of tachyonic vacua which are
classically stable but quantum mechanically unstable.
Interestingly, our method automatically excludes the
existence of instanton solutions for a potential with 
mass squared below the BF bound. Since the vacuum for such
a potential is already classically unstable, this result seems 
reasonable, though we should not perhaps claim it as a general 
theorem at the moment because we have not proved the complete 
generality of our method yet.

In order to understand better the reason behind the
quantum mechanical instability, we calculate 
various quantities as functions of the model parameters.
Comparing the algebraic criterion with our findings, we discuss 
stability criteria for tachyonic vacua in AdS spacetime. 

The paper is organized as follows. 
In section II, we formulate instanton solutions in an asymptotically
AdS spacetime.
In section III, we shows that instantons exist only for the potential 
with the mass squared above the BF bound. In section IV, 
we find exact instanton solutions, which represent the instability of 
the tachyonic state at or above the BF bound.
The final section is devoted to discussion and conclusion.

\section{Formalism}

We consider the $d$-dimensional Euclidean action for a scalar
 field $\phi$ coupled with gravity:
\begin{eqnarray}
S_{E} =-\frac{1}{2\kappa^2}\int_{M} d^dx\sqrt{g}~R
- \frac{1}{\kappa^2}\int_{\partial M} d^{d-1}x \sqrt{h}~K
+\int_{M} d^dx \sqrt{g}\left[~
\frac{1}{2}g^{\mu\nu}\partial_\mu\phi\partial_\nu\phi
+V(\phi)
~\right]\,,
\label{basicaction}
\end{eqnarray}
where $\kappa^2=8\pi G$, $R$ is the Ricci scalar of the metric 
$g_{\mu\nu}$, $h$ is the determinant of the induced metric on 
the boundary, and $K$ is the trace part of the extrinsic curvature.
The second term is necessary to make the variational principle 
consistent when the spacetime is non-compact,
called the Gibbons-Hawking boundary term~\cite{Gibbons:1976ue,Hawking:1995fd}.

Here we note that the critical case $m^2 =m_{\rm BF}^2$ needs
special care. In this case, one has two different theories
depending on the choice of the asymptotic boundary 
condition~\cite{Klebanov:1999tb}. One may stick to 
the original action~(\ref{basicaction}). But in this case 
it is known that the theory is unstable~\cite{Hertog:2003xg}.
So to discuss the quantum instability, we consider
the theory with an additional boundary term,
\begin{eqnarray}
 S_{\rm B} 
= - \frac{1}{2}\int_{\partial M} d^{d-1}x
  \sqrt{h} n^\mu \phi \partial_\mu \phi\,,
\label{extra}
\end{eqnarray}
where $n^\mu$ denotes a unit normal vector of the boundary $\partial  M$.
It is known that adding this boundary term makes the system perturbatively 
stable~\cite{Klebanov:1999tb}. In subsequent sections we
consider this theory when the mass squared saturates the BF bound. 

Assuming $O(d)$-symmetry, we consider the metric of the form,
\begin{eqnarray}
ds^2=a(z)^2\left(dz^2 + d\Omega^2_{d-1}\right)  \,, 
\end{eqnarray}
and the scalar $\phi=\phi(z)$.
Under the $O(d)$-symmetry, the action reduces to
\begin{eqnarray}
S_E=v_{S^{d-1}}\left[-\frac{(d-1)(d-2)}{2\kappa^2}\int dz
\left(a^{d-4}\dot a^2+a^{d-2}\,\right)
+ \int dz\,a^{d-1}
\left(\frac{1}{2a}\dot\phi^2 + a\,V\right)\right]\,,
\label{action}
\end{eqnarray}
where the dot denotes a derivative with respect to 
$z$ ($\dot{~}=d/dz$)
and $v_{S^{d-1}}$ is the volume of a unit 
$(d-1)-$dimensional sphere.
The equations of motion are
\begin{eqnarray}
(d-1)(d-2)\left(\frac{\ddot a}{a}-1\right)
=\kappa^2\left[-(d-3)\dot\phi^2-4a^2V\right]\,,
\label{metric1}
\end{eqnarray}
and 
\begin{eqnarray}
\ddot\phi+(d-2)\frac{\dot a}{a}\dot\phi-a^2\frac{dV}{d\phi}=0  \, .
\label{sc1}
\end{eqnarray}
The Hamiltonian constraint, which is an integral of
 (\ref{metric1}) with a specific integration constant, is
\begin{eqnarray}
(d-1)(d-2)\left[\left(\frac{\dot a}{a}\right)^2-1\right]
=2\kappa^2\left(\frac{1}{2}\dot\phi^2-a^2V\right)\, .
\label{hc1}
\end{eqnarray}

We now construct an exact instanton solution in the presence 
of gravity by extending the method developed 
in \cite{Kanno:2011vm,Kanno:2012zf}.
Namely, instead of giving the form of the potential first, we
consider the condition on the form of the scale factor for the
existence of a regular instanton solution and look for 
a function describing the scale factor that enables us to 
derive the potential as a function of the scalar field analytically.

Since we are interested in the decay of an AdS vacuum,
instantons are required to be asymptotically AdS.
For a pure AdS spacetime with the curvature radius $\ell$,
we have $a=\ell/\sinh z$ ($0<z<\infty$) where $z\to0$ 
corresponds to spatial infinity and $z\to\infty$ to the origin.
Thus for an asymptotically AdS spacetime,
we set the scale factor in the form,
\begin{eqnarray}
 a(z) = \frac{\ell}{\sinh z} ~f(\tanh z ) \ , 
\qquad\quad \lim_{z\rightarrow 0} f = 1 \ , 
\end{eqnarray}
where the function $f(\tanh z)$ is assumed to be regular at 
$0<z<\infty$, and $\ell$ is the AdS curvature radius
in the asymptotic region. Since the function $f$ describes
deformation from AdS spacetime, we call $f$ the deformation
function.
The above form of the scale factor guarantees the asymptotic 
AdS nature of the metric.
For convenience, we introduce the variable $x$ by $x=\tanh z$.
In terms of $x$, we have
\begin{eqnarray}
e^z=\sqrt{\frac{1+x}{1-x}}\,,\qquad
\frac{d}{dz}=(1-x^2)\frac{d}{dx}\,,
\qquad
\frac{d}{dx}=\cosh^2z\frac{d}{dz}\,,
\end{eqnarray}
and the scale factor is expressed as
\begin{eqnarray}
a=\ell~\frac{\sqrt{1-x^2}}{x}\,f(x)\,.
\end{eqnarray}
The range $0<z<\infty$ is mapped onto $0<x<1$.

Using Eqs.~(\ref{metric1}) and (\ref{hc1}), we can express
$\dot\phi^2$ and $V$ in terms of $\ddot a/a$ and $\dot a/a$, and 
hence in terms of the function $f(x)$ and its derivatives.
The resulting expressions are
\begin{eqnarray}
\frac{\kappa^2}{2(d-2)} \left( \frac{d\phi}{dx} \right)^2 
= \left( \frac{f'}{f} \right)^2 - \frac{1}{2} \frac{f''}{f}
 - \frac{1}{x}\frac{f'}{f}\,,
\label{dphi}
\end{eqnarray}
and
\begin{eqnarray}
\frac{2\kappa^2 \ell^2}{d-2}\,V = 
- \frac{x^2 (1-x^2)}{f^2} \left\{\frac{f''}{f}
 + (d-3)\left(\frac{f'}{f}\right)^2 
\right\} + 2 \frac{x(x^2 + d-2)}{f^2} \frac{f'}{f}
  - \frac{d-1}{f^2}  \,,
\label{potential}
\end{eqnarray}
where the prime denotes an $x$-derivative (${}'=d/dx$).
Thus, if we specify the function $f(x)$ in these 
equations, both of $d\phi/dx$ and $V$ are given as
a function of $x$. Integrating the expression for $d\phi/dx$
one obtains $\phi$ as a function of $x$. Then combining
this with the expression for $V$ as a function of $x$,
we obtain $V$ as a function of $\phi$.
In this way, we obtain an instanton solution
$\phi=\phi(x)$ for the potential $V=V(\phi)$.
In subsequent sections, we look for instanton solutions for
potentials that have tachyonic vacua.

Once the function $f$ is specified, the action
for the instanton can be easily computed.
Substituting the Hamiltonian constraint (\ref{hc1})
 into the action (\ref{action}), we manipulate
\begin{eqnarray}
S_{\rm instanton}
& =& 2v_{S^{d-1}} \int_{0}^{\infty}dz
 \left[ a^d V - \frac{(d-1)(d-2)}{2\kappa^2}\,a^{d-2}  \right]
\cr
&=&\frac{(d-2)\ell^{d-2}}{\kappa^2}v_{S^{d-1}}
\int_{0}^{1}dx\,\frac{(1-x^2)^{\frac{d}{2}-1}}{x^d}f^{d-2}
\cr
&&\hspace{1cm}\times\left[
-x^2(1-x^2)\left\{\frac{f''}{f}+(d-3)\left(\frac{f'}{f}\right)^2
\right\}
+2x(x^2+d-2)\frac{f'}{f}
-\frac{d-1}{1-x^2}
\,\right]\,.
\label{actionwhc}
\end{eqnarray}
Note that we add the boundary term (\ref{extra}),
\begin{eqnarray}
S_B=\frac{v_{S^{d-1}}\ell^{d-2}}{2}
\left[\frac{1}{x^{d-2}}\phi\frac{d\phi}{dx}\right]_{x\to0}\,,
\label{BFbterm}
\end{eqnarray}
to the above in the case the mass squared saturates the BF bound.

For an asymptotically AdS instanton solution, 
the above action (\ref{actionwhc}) diverges. However, for 
instantons that contribute to the vacuum decay,
the difference between this action and the AdS action
$S_{\rm AdS}$ should be finite, where $S_{\rm AdS}$ is given by
\begin{eqnarray}
S_{\rm AdS}=\frac{(d-2)\ell^{d-2}}{\kappa^2}v_{S^{d-1}}
\int_{0}^{1}dx\,\frac{(1-x^2)^{\frac{d}{2}-1}}{x^d}
\left[-\frac{d-1}{1-x^2}\right]\,.
\end{eqnarray}
The decay rate is given by $\Gamma\sim e^{-B}$ with 
\begin{eqnarray}
B &=& S_{\rm instanton} - S_{\rm AdS}
\cr
  &=& \frac{(d-2)\ell^{d-2}}{\kappa^2}v_{S^{d-1}} \int^1_0 dx
\frac{(1-x^2)^{\frac{d}{2}-1}}{x^d} 
\cr
&&\qquad\times\left[\frac{d-1}{1-x^2}(1-f^{d-2})
 + 2x(x^2 +d-2) f^{d-3}f' -x^2 (1-x^2) f^{d-2} 
\left\{ \frac{f''}{f} + (d-3)\left(\frac{f'}{f}\right)^2 
\right\}
\right]\,.
\label{Bdef}
\end{eqnarray}
In the case $m^2=m_{\rm BF}^2$, we add (\ref{BFbterm}) to the above.

The condition that the above integral be finite
constrains the behavior of the deformation function $f$ 
in the asymptotically AdS region.
Namely, the solution should approach the AdS spacetime
sufficiently fast to make the integral converge.
 Analyzing the behavior of the integrand of Eq.~(\ref{Bdef})
at $x=0$, which we perform in the next section,
we find that the deformation function should satisfy
\begin{eqnarray}
1-f  =O\left( x^{n} \right) \quad
\mbox{where}\quad
n=d-2\quad\mbox{or}\quad n\geq d-1  \, .
\label{condition}
\end{eqnarray}

\section{Existence of instantons satisfying the BF bound}
\label{sec:analysis}

Here we consider the condition for an instanton to exist 
based on our method presented in the previous section.
Since we are interested in the role of the BF bound
in the quantum stability, we seek for instantons describing 
the decay of a tachyonic state.
In this respect we note that instantons we look for resemble
the Linde-Lee-Weinberg instanton~\cite{Linde:1981zj,Lee:1985uv} 
in the flat spacetime, in that there is no potential barrier.

Let us first perform the asymptotic analysis
of an instanton solution in the asymptotically AdS region.
We assume that the deformation function can be expanded 
around $x\sim 0$ as
\begin{eqnarray}
  f = 1 - b\, x^n + \cdots  \ ,
\label{expansion}
\end{eqnarray}
where $b$ is a constant and $n>0$.
As long as we restrict ourselves to $O(d)$ symmetric
solutions, this assumption seems quite general. 

For $f$ of the form (\ref{expansion}), Eq.~(\ref{dphi}) gives
\begin{eqnarray}
\frac{\kappa^2}{2(d-2)}\left( \frac{d\phi}{dx} \right)^2 
= \frac{n(n+1)}{2}\, b\, x^{n-2}  +\cdots\,.
\end{eqnarray}
Since the left hand side of the above equation 
is positive, $b$ has to be positive. 
Then it is can be solved as
\begin{eqnarray}
\phi = \frac{2}{\kappa} 
\sqrt{\frac{(d-2)(n+1)}{n}\,b}\ x^{n/2} + \cdots\,.
\label{app:scalar}
\end{eqnarray}

On the other hand, Eq.~(\ref{potential}) gives the potential
\begin{eqnarray}
\kappa^2 \ell^2 V = -\frac{(d-1)(d-2)}{2}  
       + \frac{n(n-2d+2)}{8} \kappa^2 \phi^2 + \cdots \ .
\end{eqnarray}
where we have used Eq.~(\ref{app:scalar}). 
The mass of this system is read off from the coefficient of $\phi^2$,
\begin{eqnarray}
m^2 \ell^2 &=&  \frac{n(n-2d+2)}{4} 
\cr
&=& \frac{\left( n- d+1 \right)^2}{4} +m_{\rm BF}^2\ell^2 \ .
\label{mass1}
\end{eqnarray}
where $m^2_{\rm BF}$ is defined in Eq.~(\ref{bf}). 
This tells us that the mass squared always satisfies the BF bound
and it is tachyonic for $n<2d-2$ and 
saturates the BF bound when $n= d-1$. 

Note that we can express $n$ in terms of $m^2$ by 
inverting the above relation,
\begin{eqnarray}
  \frac{n}{2} = \frac{d-1 \pm \sqrt{(d-1)^2 + 4m^2 \ell^2}}{2} \,.
\label{npm}
\end{eqnarray}
This agrees with the standard asymptotic behavior of a
scalar field with mass squared $m^2$ in asymptotically AdS spacetime.

For an instanton to mediate the vacuum decay, the
decay rate should be finite. This can be examined by
expanding the integrand of the bounce action $B$ given by
Eq.~(\ref{Bdef}) around $x=0$. The result is
\begin{eqnarray}
  B = \frac{(d-2)\ell^{d-2}\,v_{S^{d-1}}}{\kappa^2} \int^1_0 
\frac{dx}{x^d} \left[   (n-d+1)(n-d+2)\, b \,x^n + \cdots \right]
\ .
\end{eqnarray}
The decay rate is apparently finite for $n>d-1$, that is, 
$m^2 > m_{\rm BF}^2$. 
In the critical case of $n=d-1$, the coefficient
of $x^n$, which would lead to a logarithmic divergence, vanishes. 
Hence, the action converges even in this critical case. 
In addition, the boundary term (\ref{BFbterm}) is evaluated as
\begin{eqnarray}
S_B=  \frac{(d-2)\ell^{d-2}\,v_{S^{d-1}}}{\kappa^2}\,
 b\,(n+1)x^{n-d+1}|_{x\to0}\,.
\end{eqnarray}
Since the critical case corresponds to $n=d-1$, this is indeed
finite and contributes to the bounce action. Note that
the boundary term vanishes for $n>d-1$. 

Interestingly there is yet another case for which $B$ is finite.
For $n=d-2$, one also finds the leading order coefficient
vanishes. By analyzing the next leading order term, one further
finds it vanishes as well. Hence $B$ is finite in this case, too.

To summarize the decay rate is finite for $n=d-2$ and $n\geq d-1$.
Note that the above analysis is quite general in the sense
that it involves only the asymptotic behavior in the asymptotically
AdS region and we have not imposed any condition on the mass
of the scalar field a priori. 
We have only assumed the existence of an instanton with 
the asymptotic behavior~(\ref{expansion}) and imposed
the positivity of $(d\phi/dx)^2$. 
Therefore, it should be applicable to any Euclidean solutions.
The important conclusion at this stage is that 
if there exists an instanton that satisfies
the asymptotic behavior~(\ref{expansion}), it must have 
its mass squared at or above the BF bound.

\section{Exact Instanton Solutions}

In the previous section, we have shown that instantons exist only 
for potentials satisfying the BF bound. 
In this section, we explicitly find a series of
exact instanton solutions.

We consider a solution given by
\begin{eqnarray}
 f(x) = \frac{c}{c+x^n}
\qquad(c>0)\,.
\end{eqnarray}
where $n$ is an arbitrary real number satisfying $n=d-2$
or $n\geq d-1$. 
We notice that the asymptotic form,
\begin{eqnarray}
 f(x) \sim 1 - \frac{1}{c} x^n + \cdots 
\end{eqnarray}
matches the expansion (\ref{expansion}).
The parameter $n$ determines the mass squared at the metastable
vacuum as seen from the analysis in the previous section,
while $c$ determines the global shape of the potential.

Inserting the above into (\ref{dphi}), we find
\begin{eqnarray}
\kappa \frac{d\phi}{dx} 
  = \pm\sqrt{\frac{(d-2)n(n+1) x^{n-2} }{c +x^n}} \ .
\end{eqnarray}
This can be readily integrated to give
\begin{eqnarray}
\kappa\phi =
 \pm\sqrt{\frac{4(d-2)(n+1)}{n}} \sinh^{-1} \frac{x^{n/2}}{\sqrt{c}} \,,
\qquad
x^{n/2} =\sqrt{c}\sinh\left(\kappa\sqrt{\frac{n}{4(d-2)(n+1)}}|\phi|
\right)\,.
\label{xphi}
\end{eqnarray}
Since $0<x<1$, the range of the scalar field is restricted to
$0<\phi<\sqrt{\frac{4(d-2)(n+1)}{n\kappa^2}}\sinh^{-1}1/\sqrt{c}$.
The instanton covers this range of the potential.

\begin{center}
\begin{figure}[t]
\includegraphics[width=95mm]{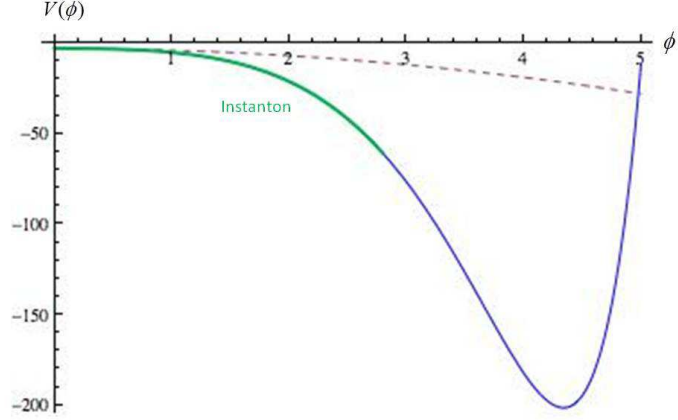}
\caption{ 
The potential as a function of $\phi$ is depicted for $c=1$ and $n=4$.
The green line is the range the instanton runs. The dashed 
line shows the potential $V=-3/(\kappa\ell)^2-m^2\phi^2/2$ for
comparison. The horizontal axis is in units of $\kappa\phi$
and the vertical axis in units of $1/(\kappa\ell)^2$.
}
\label{fig:1}
\end{figure}
\end{center}

From (\ref{potential}), the potential is given by
\begin{eqnarray}
\frac{2\kappa^2 \ell^2}{d-2} V &=& -(d-1) +\frac{(n+1)(n-2d+2)}{c}~x^n 
- \frac{n(n+1)}{c}~x^{n+2} 
\cr &&\qquad
- \frac{(n+1)\{(d-2)n+d-1\}}{c^2}~x^{2n}
+ \frac{n\{(d-2)n-1\}}{c^2}~x^{2n+2}\ .
\end{eqnarray}
Inserting (\ref{xphi}) into the above, we obtain
the potential as a function of $\phi$.
\begin{eqnarray}
\frac{2\kappa^2 \ell^2}{d-2}  V &=& -(d-1) + (n+1)(n-2d+2) \sinh^2 
\left(\kappa\sqrt{\frac{n}{4(d-2)(n+1)}}\,|\phi|\right) 
\cr
&&
- n(n+1)~c^{\frac{2}{n}} \sinh^{2+\frac{4}{n}} 
\left(\kappa\sqrt{\frac{n}{4(d-2)(n+1)}}\,|\phi|\right) 
\cr
&&
- (n+1)\{(d-2)n+d-1\} \sinh^4 
\left(\kappa\sqrt{\frac{n}{4(d-2)(n+1)}}\,|\phi|\right) 
\cr
&&
+ n\{(d-2)n-1\}~c^{\frac{2}{n}} \sinh^{4+\frac{4}{n}} 
\left(\kappa\sqrt{\frac{n}{4(d-2)(n+1)}}\,|\phi|\right)  \ .
\end{eqnarray}
Note that there is a branch cut at $\phi=0$ except for $n= 1$, $2$ and $4$.
However, the potential is sufficiently smooth for the other values
of $n$ in the sense that its second derivative is finite and
continuous at $\phi=0$.
As an example, the potential in the case of $d=4$, $n=4$ and $c=1$ 
is shown in Fig.~\ref{fig:1}.

If we expand the above potential around $\phi=0$, we find
\begin{eqnarray}
&&V= -\frac{(d-1)(d-2)}{2\kappa^2\ell^2}
 +\frac{m^2}{2}\phi^2 + \cdots\,;
\cr
\cr
&&\quad
m^2 = \frac{n(n-2d+2)}{4\,\ell^2}
=\frac{\left( n- d+1 \right)^2}{4\,\ell^2} +m_{\rm BF}^2\,,
\quad
m_{\rm BF}^2=-\frac{(d-1)^2}{4\,\ell^2}\,.
\end{eqnarray}
We see that the mass squared agrees with 
the general analysis in Eq.~(\ref{mass1}). 
Thus, our instanton solutions indeed satisfy the BF bound,
implying that tachyonic vacua which are perturbatively stable 
can be quantum mechanically unstable. 

To further support this picture, we have calculated the energy at
the nucleation surface and found it is zero for $n\geq d-1$,
which is consistent with the interpretation that the analytic
 continuation of the instanton describes the state after the decay
of the AdS vacuum.
In the critical case $m^2 =m_{\rm BF}^2$, the contribution of
the boundary term (\ref{extra}) is essential to make the energy zero.
Without it the energy turns out to be negative. But this is consistent 
with the fact that the system is unstable
unless the boundary term is added.

In the case $n=d-2$, the energy is divergent to minus infinity
in spite of the fact that the bounce action $B$ is finite.
Mathematically this is because this solution corresponds to 
the minus sign of Eq.~(\ref{npm}) for $m^2\ell^2=-d(d-2)/4$, 
hence picks up the singular behavior in the asymptotically AdS region.
On the other hand, if we adopt the boundary term  (\ref{extra}),
the energy vanishes. In this case, the bounce action diverges 
to plus infinity, resulting in the vanishing decay rate.
This probably means that the instanton solution in this case 
is not relevant for the vacuum decay.
Study of the exact meaning of this solution is left as a future issue.

As shown in the previous section, the bounce action
for the above solution is finite for $n=d-2$ and $n\geq d-1$.
To confirm this and to see how the bounce action $B$ depends
on the shape of the potential, we compute it below.
Although we can compute $B$ for any spacetime dimension numerically
if necessary (even analytically in the case of even spacetime dimensions),
we focus on the case of four dimensions ($d=4$).

\subsection{The bounce action in four dimensions}

We focus on four dimensions, $d=4$. 
As a simple example that satisfies the condition~(\ref{condition}), 
we first consider the case $n=4$,
\begin{eqnarray}
 f(x) = \frac{c}{c+x^4} \qquad(c>0)\,.
\end{eqnarray}
For this, we obtain a completely analytical potential, 
\begin{eqnarray}
 \kappa^2 \ell^2  V = -3 - 10 \sinh^2 \frac{\kappa \phi}{\sqrt{10}}
- 20 \sqrt{c} \sinh^3 \frac{\kappa \phi}{\sqrt{10}}
 - 55 \sinh^4 \frac{\kappa \phi}{\sqrt{10}}
 + 28 \sqrt{c} \sinh^5 \frac{\kappa \phi}{\sqrt{10}} \ .
\label{typical}
\end{eqnarray}
This potential is shown in Fig.~\ref{fig:1} for the parameter $c=1$.
The bounce action~(\ref{Bdef}) can be also evaluated analytically as
\begin{eqnarray}
B=B_4&\equiv& \frac{\pi^2 \ell^2}{8\kappa^2}
 \left[ \frac{8}{1 + c} 
- \frac{10 \sqrt{2}}{c^{1/4}}
 \tan^{-1} \left(1 - \frac{\sqrt{2}}{c^{1/4}}\right) 
+ \frac{10 \sqrt{2}}{c^{1/4}} 
\tan^{-1} \left(1 + \frac{\sqrt{2}}{c^{1/4}}\right) 
+ \frac{5 \sqrt{2}}{c^{1/4}} 
\log \frac{1 - \sqrt{2} c^{1/4} + \sqrt{c}}
{1 + \sqrt{2} c^{1/4} + \sqrt{c}} \right]
\cr
\cr
&=&\frac{\ell^2}{\kappa^2}\times
\left\{
\begin{array}{ll}
\displaystyle\frac{5\sqrt{2}\,\pi^3}{4\,c^{1/4}}
-4\pi^2+O(c^2)
\quad &\mbox{for}\,\ c\ll1\,,
\\
~
\\
\displaystyle\frac{8\pi^2}{3\,c}
\left(1-\frac{9}{14c}+O(c^{-2})\right)
\quad &\mbox{for}\,\ c\gg1\,,
\end{array}\right.
\label{B:n=4}
\end{eqnarray}
where we used $v_{S^{3}} = 2\pi^2$.

\begin{figure}[ht]
\begin{center}
\begin{minipage}{8.5cm}
\begin{center}
\vspace{-7mm}
\hspace{-1.5cm}
\includegraphics[width=95mm]{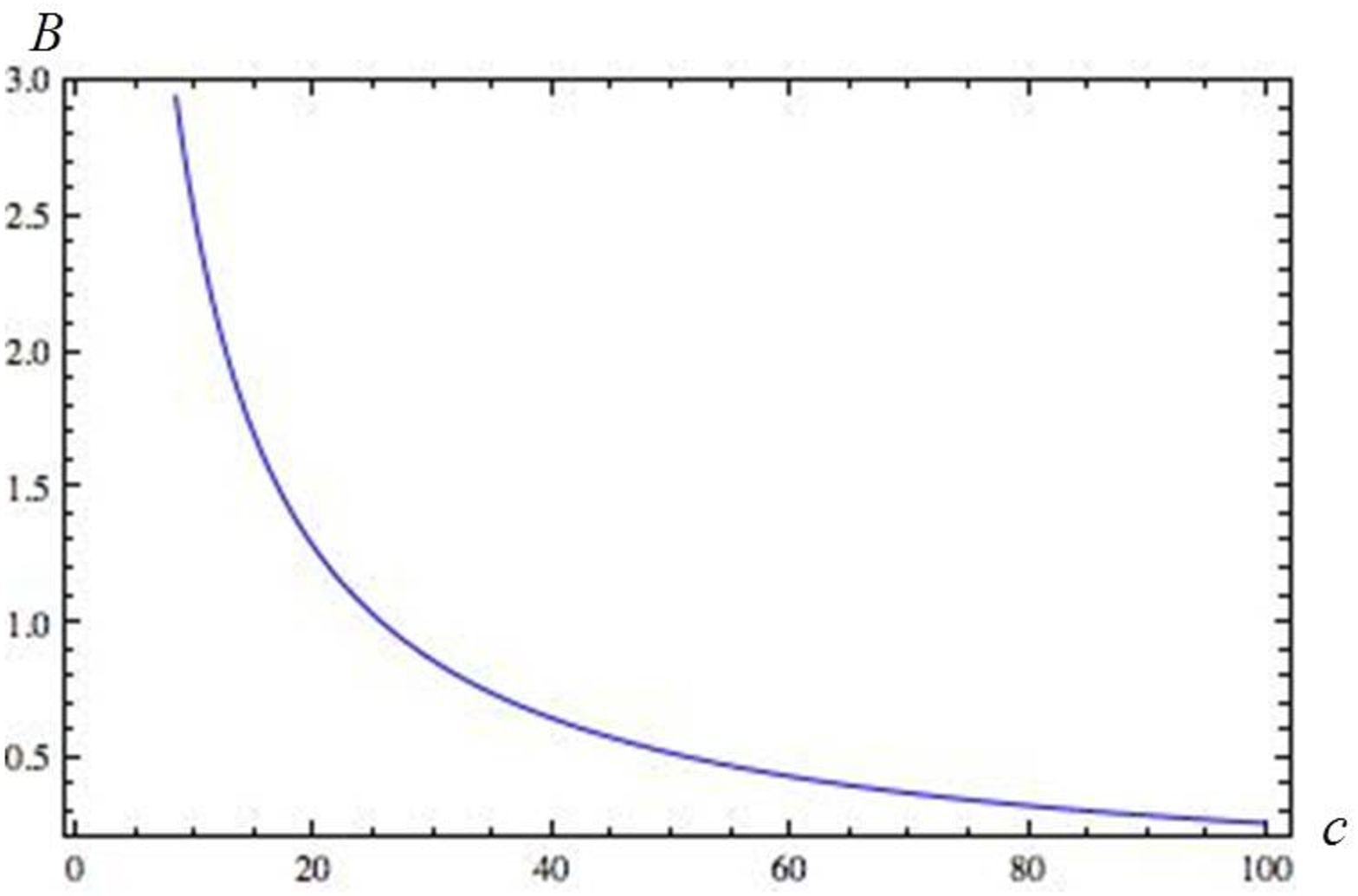}\vspace{0cm}
\caption{The bounce action as a function of $c$ is depicted for 
$n=4$. The decay rate is larger for large $c$.}
\label{fig:2}
\end{center}
\end{minipage}
\hspace{1mm}
\begin{minipage}{8.5cm}
\begin{center}
\hspace{-1.5cm}
\includegraphics[width=95mm]{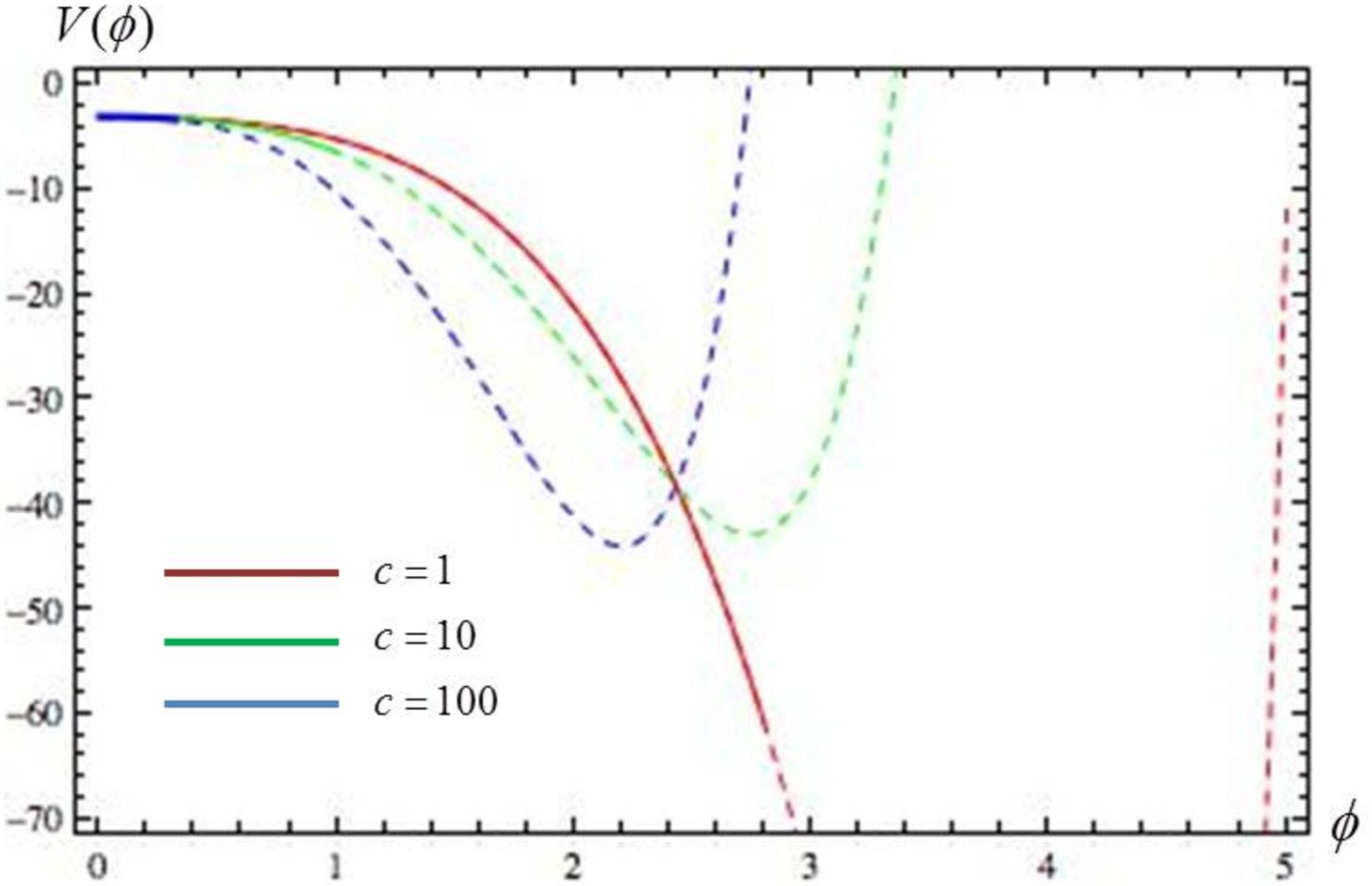}\vspace{0cm}
\caption{The potential in the case of $n=4$
as a function of $\phi$ for several values of $c$ 
(drawn with the dashed lines). The units are the
same as in Fig.~1.
The real lines shows the region that instanton run.}
\label{fig:3}
\end{center}
\end{minipage}
\end{center}
\end{figure}

Let us now turn to general cases of arbitrary $n$,
\begin{eqnarray}
 f(x) = \frac{c}{c+x^n}   \qquad(c>0)\,.
\end{eqnarray}
For this instanton, for $n=2$ or $n > 3$,
we can calculate the bounce action as
\begin{eqnarray}
B_n &=&\frac{4\pi^2 \ell^2 }{\kappa^2}
\left[-1-\frac{c}{(1+c)n}
+\frac{1+n}{n}
F\left( 1,\,- \frac{1}{n},\,\frac{n-1}{n} ;\, -\frac{1}{c}\right)
\right]
\cr
\cr
&=& \frac{\ell^2}{\kappa^2}\times
\left\{
\begin{array}{ll}
\displaystyle
\frac{4(n+1)\pi^3}{n^2\sin(\pi/n)\,c^{1/n}}-4\pi^2+O(c)
\quad &\mbox{for}\,\ c\ll1\,,
\\
~\\
\displaystyle
\frac{8\pi^2}{n-1}\frac{1}{c}
\left(1+O(c^{-1})\right)
\quad &\mbox{for}\,\ c\gg1\,,
\end{array}\right.
\label{B:n}
\end{eqnarray}
where $F(\alpha,\,\beta,\,\gamma\,;\,u)$ is a hypergeometric function.
We see the bounce action decreases as $c$ or $n$ increases.

In the case $n=3$ which corresponds to the BF bound $m^2 = m_{\rm BF}^2$,
we add the boundary term (\ref{BFbterm}) to obtain
\begin{eqnarray}
B_3 &=&\frac{4\pi^2 \ell^2 }{\kappa^2}
\left[-1-\frac{c}{(1+c)n}
+\frac{1+n}{n}
F\left( 1,\,- \frac{1}{n},\,\frac{n-1}{n} ;\, -\frac{1}{c}\right)
- \frac{1}{c}
\right] \Biggl|_{n=3}  +  S_{\rm B} \nonumber \\
&=& \frac{4\pi^2 \ell^2 }{\kappa^2}
\left[-1-\frac{c}{(1+c)3}
+\frac{1+3}{3}
F\left( 1,\,- \frac{1}{3},\,\frac{2}{3} ;\, -\frac{1}{c}\right)
 +\frac{3}{c} \right]
 \nonumber\\
&=& \frac{4\pi^2\ell^2}{\kappa^2}\times 
\left\{
\begin{array}{ll}
\displaystyle
\frac{3}{c}+\frac{8\sqrt{3}\pi}{27\,c^{1/3}}-1+O(c)
\quad &\mbox{for}\,\ c\ll1\,,
\\
~\\
\displaystyle
\frac{4}{c}
\left(1+O(c^{-1})\right)
\quad &\mbox{for}\,\ c\gg1\,,
\end{array}\right.
\end{eqnarray}
where we have used $S_B =  16 \pi^2 \ell^2 /(\kappa^2 c)$.
Notice that we have an extra contribution $-1/c$ from the bulk
integral when $n=3$. Because of this extra contribution the bounce
action would become negative if the boundary term were not added.
Again, this seemingly pathological result is consistent with the fact 
that the system would be already perturbatively unstable,
implying that the instability would develop classically without
any barriers, if the boundary term were not added.

In Fig.~\ref{fig:2}, we plot $B$ as a function of $c$. 
As expected from the above equation the decay rate 
is larger large for larger $c$. 
This means increasing $c$ makes the tunneling process easier,
or renders the system more unstable.
Since there is no dependence of $c$ in the mass, 
changing $c$ implies the changing the non-linear terms of 
the potential. To see the actual shape of the potential, 
we plot it in Fig.~\ref{fig:3}.
 Apparently, the potential gets steeper near the origin 
$\phi=0$ and the region of the instanton runs shrinks 
as we increase $c$.  Thus, we see that the larger $c$ 
makes the potential more unstable. This explains the
tendency that the decay rate is larger for larger $c$.

\begin{figure}[ht]
\begin{center}
\begin{minipage}{8.5cm}
\begin{center}
\vspace{-7mm}
\hspace{-1.5cm}
\includegraphics[width=95mm]{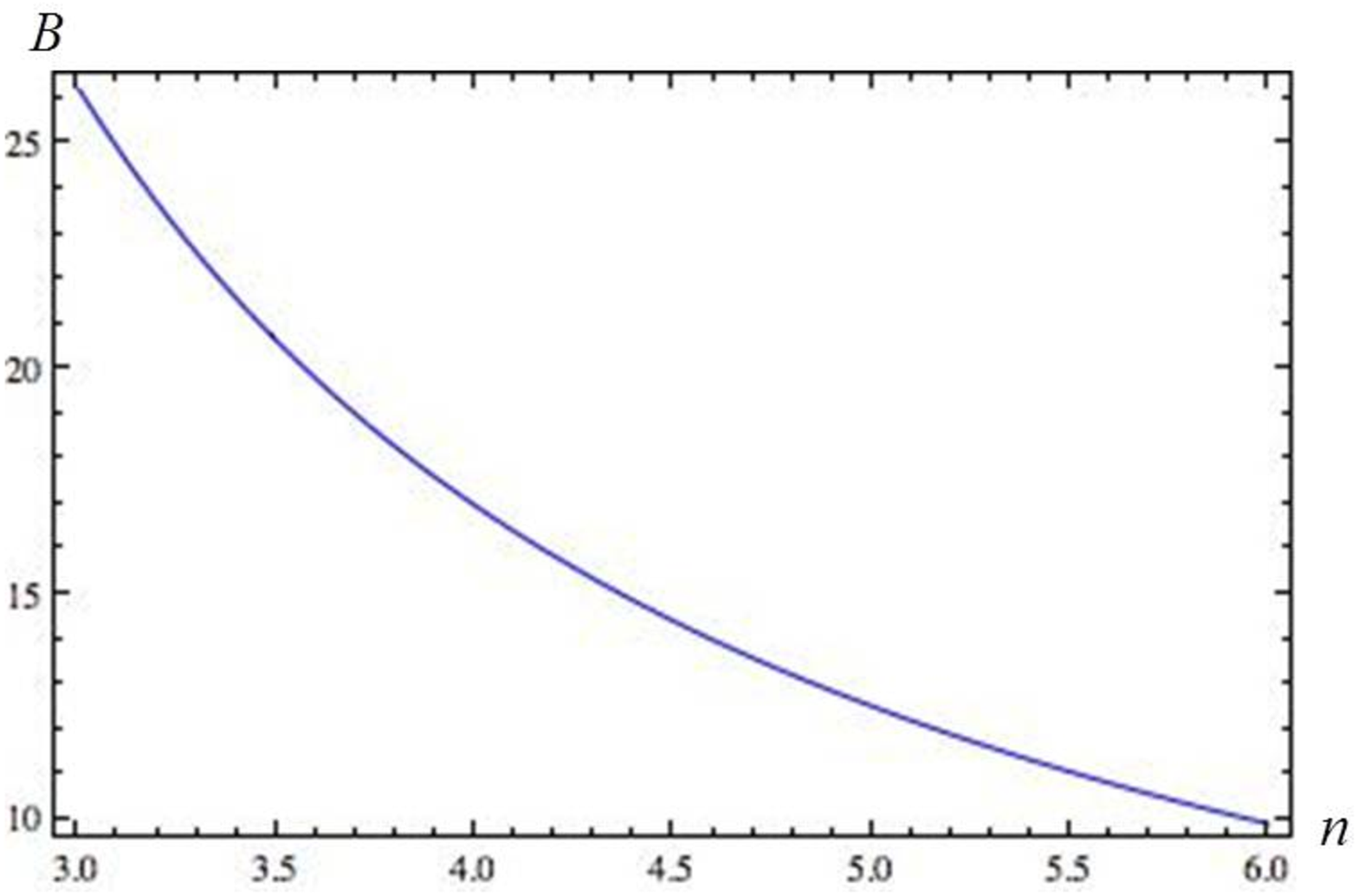}\vspace{0cm}
\caption{The bounce action as a function of $n$ is 
depicted for $c=1$. This shows that tunneling is easier for
large $n$.
}
\label{fig:4}
\end{center}
\end{minipage}
\hspace{1mm}
\begin{minipage}{8.5cm}
\begin{center}
\hspace{-1.5cm}
\includegraphics[width=95mm]{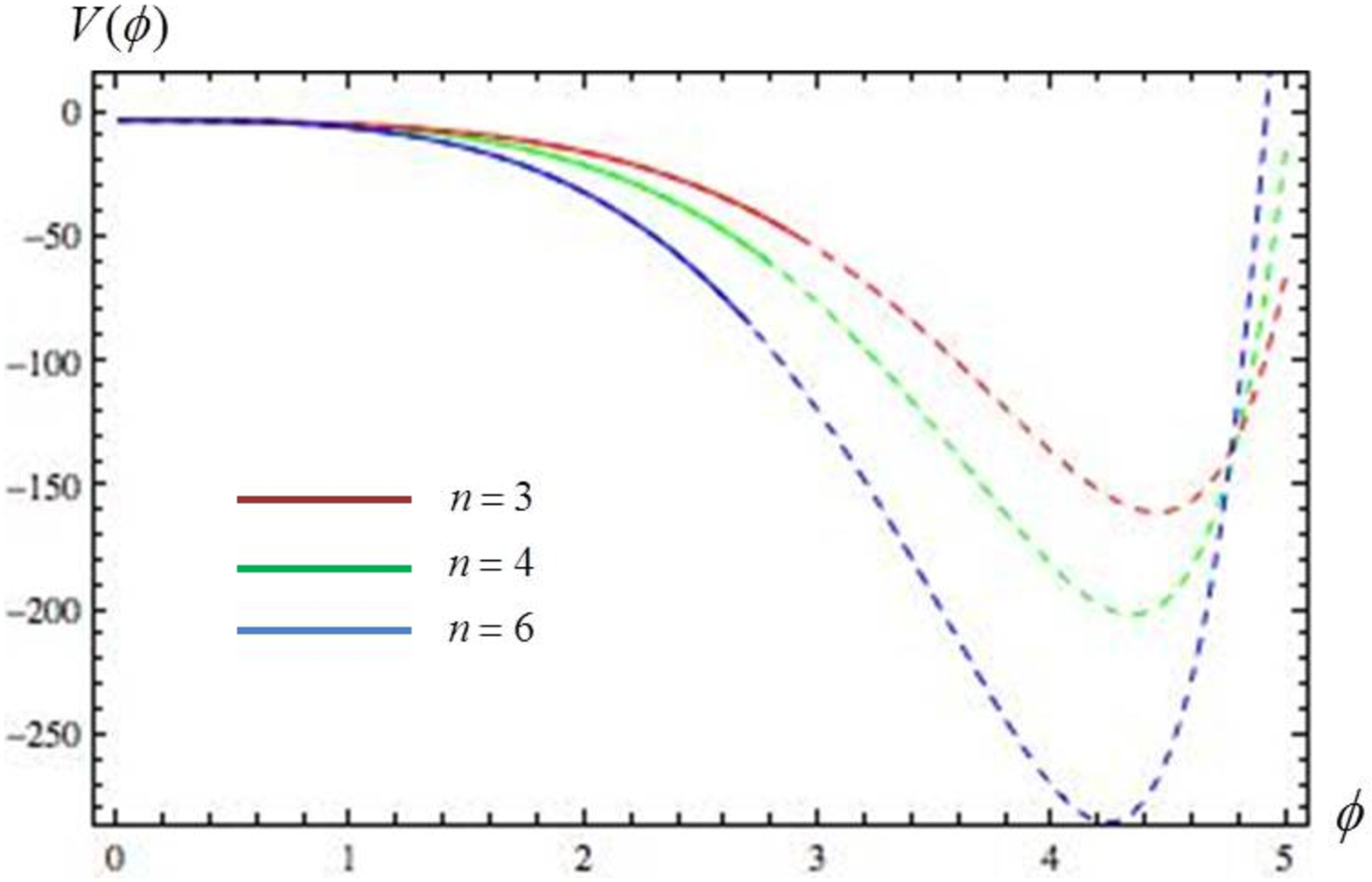}\vspace{0cm}
\caption{The potential for $c=1$ as a function of $\phi$ for 
several different values of $n$ (drawn with the dashed lines). 
The units are the same as in Fig.~1.
The real lines shows the region that instanton run.}
\label{fig:5}
\end{center}
\end{minipage}
\end{center}
\end{figure}

In Fig.~\ref{fig:4}, we show the bounce action $B$ as a function of $n$.
It clearly shows that $B$ decreases as $n$ increases.
It should be noted that for $n>6$ the mass squared is positive 
and hence there appears a potential barrier. 
This fact seems counter-intuitive because the tunneling becomes
easier to occur as the barrier grows higher.
However, changing $n$ also changes the nonlinear part of the potential.
So we need to see how the nonlinear part of the potential
depends on $n$. In Fig.~\ref{fig:5}, potentials for several values of
 $n$ are shown. We see that the potential becomes deeper as $n$ 
increases. Thus in spite of the increase of the mass squared at
the origin, the potential at larger values of $\phi$
becomes deeper for larger $n$ and the system becomes more unstable.

\subsection{Comparison with stable potentials}

\begin{center}
\begin{figure}
\includegraphics[width=115mm]{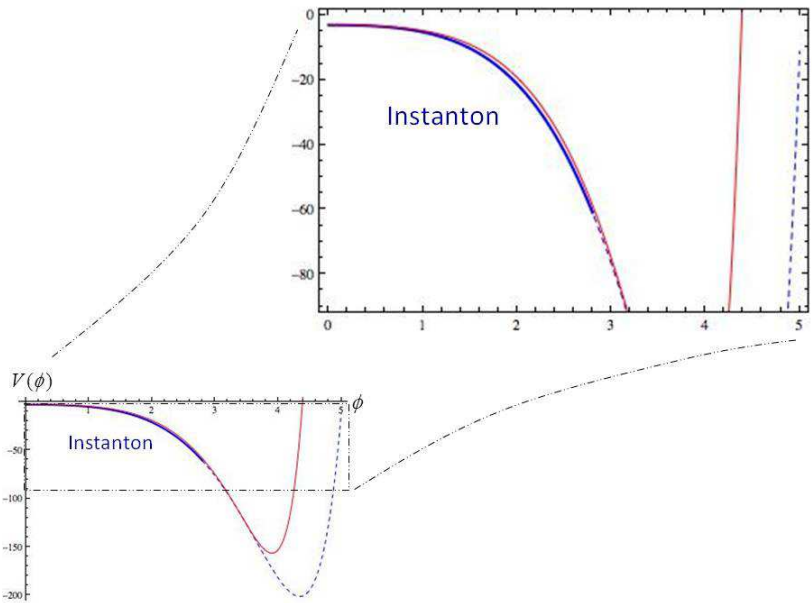}
\caption{Comparison of stable and unstable potentials.
The red curve shows a stable potential, the blue dashed curve
shows an unstable potential which behaves almost identically with
the stable potential near the origin. 
The thick blue line shows the region that the instanton covers.}
\label{fig:6}
\end{figure}
\end{center}

We have obtained a series of quantum mechanically unstable potentials.
We can also generate a stable potentials by inspection 
using an algebraic condition (\ref{super}).
For example, if we take
\begin{eqnarray}
P(\phi) = 1+ \frac{1}{2}\phi^2 + \frac{1}{300}\phi^3
 - \frac{1}{4000}\phi^4+ \frac{1}{90000}\phi^6  \ ,
\end{eqnarray}
we obtain the potential,
\begin{eqnarray}
V(\phi) &=& -3 - \phi^2 +\frac{3}{50} \phi^3 
- \frac{7507}{10000}\phi^4 -\frac{753}{25000}\phi^5
+\frac{163}{250000}\phi^6  \nonumber\\
&& \qquad + \frac{23}{1000000}\phi^7  -\frac{2703}{80000000}\phi^8
-\frac{1}{1500000}\phi^9 + \frac{23}{900000000}\phi^{10}
 - \frac{1}{2700000000}\phi^{12} \ .
\end{eqnarray}
In Fig.~\ref{fig:6}, we compared the above potential with 
the potential (\ref{typical}) with $c=1$.
 Taking a close look at the potential,
we find the unstable potential is slightly below the stable one.
This small difference is crucially important in 
the determination of the stability.
As $c$ decreases, the difference between the two potentials
near the origin becomes even smaller. Together with the fact
that $B$ diverges as $c^{-1/4}$ as seen in the second line of 
Eq.~(\ref{B:n=4}) or (\ref{B:n}), this suggests that our instanton 
solution verges towards the instability boundary for $c\ll1$.

\section{Conclusion}

It is known that in the asymptotically AdS spacetime,
a tachyonic vacuum is perturbative stable if the mass squared is at
or above the BF bound. But this does not imply the global stability.
To discuss it one may resort to a different 
criterion~\cite{Boucher:1984yx,Townsend:1984iu}. Namely, 
if the potential is written as
$$
  V(\phi) = 
(d-2) \left(\frac{dP(\phi)}{d\phi}\right)^2 - (d-1) P(\phi)^2\,,
$$
with some function $P(\phi)$, the vacuum is stable. 
This condition is a sufficient condition for the stability.
The above equation may be viewed as a nonlinear differential equation
for $P$ given the potential $V$. Unfortunately, however,
it is formidable to see if there is a real solution
for $P$ or not for a generic potential $V$. 

In this paper, we took another route, that is, we
looked at a sufficient condition for the instability
instead of stability. In other words, we looked for
instantons with finite decay rate. If we find such an instanton,
it proves the instability of the vacuum.
We indeed found a series of exact instanton solutions
for a class of potentials which satisfy the BF bound.
These instantons have finite bounce action, hence describe 
the decay of (tachyonic) vacua.
In other words, the class of potentials we found must not
be expressed in the form of the above equation.

In passing we also showed that there is no instanton solution
if the mass squared of the potential is below the BF bound,
under a physically reasonable assumption for the asymptotic
behavior of the instanton in the asymptotic AdS region.

In conclusion, we found exact instanton solutions which
 destabilize tachyonic vacua at or above the BF bound. 
We also noted that our solution can be extended to vacua with barriers
if we allow a slight violation of analyticity of the potential, 
in the sense that the potential is kept smooth up to the second 
derivative (ie it is $C^2$).

It is interesting to explore general exact CDL instantons using 
our method by relaxing the requirement of exactness if necessary. 
It is also intriguing to explore implication of our instantons 
to the AdS/CFT correspondence. 

\acknowledgements

We would like to thank Akihiro Ishibashi for useful discussions.
This work is supported in part by the JSPS
Grants-in-Aid for Scientific Research (C) No.~22540274 and (A) No.22244030,
the Grant-in-Aid for Creative Scientific Research No.~19GS0219, 
the Grant-in-Aid for Scientific Research on Innovative Area No.21111006.
SK is supported in part by grant PHY-0855447 from the National Science 
Foundation. SK would like to thank Yukawa Institute for 
Theoretical Physics (YITP) members at Kyoto University for warm
hospitality. A part of this work was done while SK was visiting YITP.


\begin{thebibliography}{99}

\bibitem{Coleman:1977py}
  S.~R.~Coleman,
  Phys.\ Rev.\  {\bf D15}, 2929-2936 (1977).

\bibitem{Callan:1977pt}
  C.~G.~Callan, Jr., S.~R.~Coleman,
  Phys.\ Rev.\  {\bf D16}, 1762-1768 (1977).
  
\bibitem{Coleman:1980aw}
  S.~R.~Coleman, F.~De Luccia,
  Phys.\ Rev.\  {\bf D21}, 3305 (1980).
   
\bibitem{Breitenlohner:1982jf} 
  P.~Breitenlohner and D.~Z.~Freedman,
  Annals Phys.\  {\bf 144}, 249 (1982).
 
\bibitem{Breitenlohner:1982bm} 
  P.~Breitenlohner and D.~Z.~Freedman,
  Phys.\ Lett.\ B {\bf 115}, 197 (1982).

\bibitem{Boucher:1984yx} 
  W.~Boucher,
  Nucl.\ Phys.\ B {\bf 242}, 282 (1984).

\bibitem{Townsend:1984iu} 
  P.~K.~Townsend,
  Phys.\ Lett.\ B {\bf 148}, 55 (1984).

\bibitem{Hertog:2005hm} 
  T.~Hertog and S.~Hollands,
  Class.\ Quant.\ Grav.\  {\bf 22}, 5323 (2005)
  [hep-th/0508181].

\bibitem{Amsel:2006uf} 
  A.~J.~Amsel and D.~Marolf,
  Phys.\ Rev.\ D {\bf 74}, 064006 (2006)
  [Erratum-ibid.\ D {\bf 75}, 029901 (2007)]
  [hep-th/0605101].

\bibitem{Amsel:2007im} 
  A.~J.~Amsel, T.~Hertog, S.~Hollands and D.~Marolf,
  Phys.\ Rev.\ D {\bf 75}, 084008 (2007)
  [Erratum-ibid.\ D {\bf 77}, 049903 (2008)]
  [hep-th/0701038].

\bibitem{Faulkner:2010fh} 
  T.~Faulkner, G.~T.~Horowitz and M.~M.~Roberts,
  Class.\ Quant.\ Grav.\  {\bf 27}, 205007 (2010)
  [arXiv:1006.2387 [hep-th]].

\bibitem{Amsel:2011km} 
  A.~J.~Amsel and M.~M.~Roberts,
  arXiv:1112.3964 [hep-th].

\bibitem{Hertog:2004rz} 
  T.~Hertog and G.~T.~Horowitz,
  JHEP {\bf 0407}, 073 (2004)
  [hep-th/0406134].

\bibitem{Hertog:2005hu} 
  T.~Hertog and G.~T.~Horowitz,
  JHEP {\bf 0504}, 005 (2005)
  [hep-th/0503071].

\bibitem{Kanno:2011vm}
  S.~Kanno and J.~Soda,
  arXiv:1111.0720 [hep-th].

\bibitem{Kanno:2012zf} 
  S.~Kanno, M.~Sasaki and J.~Soda,
  arXiv:1201.2272 [hep-th].

\bibitem{Gibbons:1976ue}
  G.~W.~Gibbons, S.~W.~Hawking,
  Phys.\ Rev.\  {\bf D15}, 2752-2756 (1977).

\bibitem{Hawking:1995fd}
  S.~W.~Hawking, G.~T.~Horowitz,
  Class.\ Quant.\ Grav.\  {\bf 13}, 1487-1498 (1996).
  [gr-qc/9501014].
  
\bibitem{Klebanov:1999tb} 
  I.~R.~Klebanov and E.~Witten,
  Nucl.\ Phys.\ B {\bf 556}, 89 (1999)
  [hep-th/9905104].

\bibitem{Hertog:2003xg} 
  T.~Hertog, G.~T.~Horowitz and K.~Maeda,
  Phys.\ Rev.\ D {\bf 69}, 105001 (2004)
  [hep-th/0310054].

\bibitem{Linde:1981zj} 
  A.~D.~Linde,
  Nucl.\ Phys.\ B {\bf 216}, 421 (1983)
  [Erratum-ibid.\ B {\bf 223}, 544 (1983)].

\bibitem{Lee:1985uv}
  K.~M.~Lee and E.~J.~Weinberg,
  Nucl.\ Phys.\  B {\bf 267}, 181 (1986).



\end{thebibliography}
\end{document}